\documentclass[preprint,pre,superscriptaddress]{revtex4}

\usepackage{graphicx}

\usepackage{amssymb,amsfonts,amsmath}

\begin{document}

%%%%%%%%%%%%%%%%%%%%%%%%%%%%%%

%% For titles, only capitalize the first letter
%% \title{Almost sharp fronts for the surface quasi-geostrophic equation}

\title{Contact process with a defect:  universal oasis, nonuniversal scaling}

\author{Zvi Miller}
\affiliation{Department of Physics, Bar-Ilan University, Ramat-Gan
52900, Israel}
\author{Nadav M. Shnerb}
\affiliation{Department of Physics, Bar-Ilan University, Ramat-Gan
52900, Israel}

\begin{abstract}
The extinction transition in the presence of a localized quenched
defect is studied numerically. When the bulk is at criticality, the
correlation length diverges and even an infinite system cannot
"decouple" from the defect. The  results presented here suggest
that,  in 1+1 dimensions, the critical exponent $\delta$ that
controls the asymptotic power-law decay depends on the strength of
the local perturbation. On the other hand, the exponent was found to
be independent of the local arrangement of the defect. In higher
dimensions the defect seems to induce a transient behavior that
decays algebraically in time.
\end{abstract}
\maketitle

\section{Introduction}

In a  recent article,  the results of a spatial-genetic study of
honeybees in the Sahara desert~\cite{honey} have  been published.
The   authors   concluded that these bees (Apis mellifera) first
arrived to the desert around 10,000 years ago, when the Sahara was a
green savannah. Although  the contemporary  Sahara  is  inhospitable
to these bees, they do  survive in many \textbf{oases} that litter
the desert. The same scenario holds  for many existing communities
of animals: their habitat is made of relatively small patches, and
the sustainability of the whole system is determined by the rates of
local extinction and recolonization. When migration among "oases" is
relatively weak, such a system is known as a \emph{metapopulation}
\cite{hansk}.

In the study of metapopulation persistence, many researchers are
using the stochastic patch occupancy models (SPOM)
\cite{etienne,moilanen1,moilanen2} as a paradigmatic framework.
These models assume that each habitat patch is either occupied or
empty, i.e., that the timescale in which the population reaches its
local carrying capacity equilibrium is relatively small. Neglecting
the details of the in-site dynamics the theory is then focused on
extinction and recolonization events. In the language of stochastic
dynamics, we are dealing with a contact process~\cite{111,107}
taking place  in heterogenous environment. While the properties of
the contact process and its generalizations are quite well
understood when it takes place on a uniform lattice, much less is
known in the presence of a quenched disorder. In this paper we are
dealing with a very simple case: a (generalized) contact process
perturbed by a local heterogeneity. Interestingly it turns out that
even a local defect may alter the behavior of the whole system at
the most important parameter regime, close to the extinction
transition.

On a homogenous lattice the contact process is known to admit an
extinction transition when the birth rate is below some
(nonuniversal) threshold. For any extinction transition on   a
homogeneous substrate and a single absorbing state, Grassberger
\cite{gras} and Janssen \cite{jan} conjectured that the microscopic
details of the stochastic process are irrelevant close to the
extinction point and the transition belongs to  the directed
percolation (DP) universality class. The basic rationale behind this
conjecture is that a spatially extended system decomposes, close to
the transition, into active and inactive zones, where after each
typical period of time there is certain probability for an active
state to die, to survive, or to infect its inactive neighbors. If
these regions are considered as lattice points on a $d$ dimensional
array, the chance of an active site to survive or to infect its
neighbors within a unit time is equivalent to the chance that a bond
exists between a lattice point at time $t$ and its neighbors in a
subsequent replica of the system at $t+1$. Accordingly, the
extinction transition happens when the bond density is exactly at
threshold for an infinite cluster in a $d+1$ dimensional system, and
the transition belongs to the directed percolation universality
class in  $d+1$ dimensions.

The Grassbrger-Janssen  conjecture has proven to be extremely
robust, and a large number of stochastic models that admit an
extinction transition were shown to belong to the DP equivalence
class if the substrate is homogenous~\cite{1,2}. It was further
shown that  spatio-temporal substrate noise (i.e.,  birth-death
rates that fluctuate in space and time with only  short range
correlations)  is an irrelevant perturbation close to the
transition, so small noise is averaged out and leaves the DP
transition unaffected~\cite{1}, although it can change the location
of the transition point~\cite{levin3}. The case of diffusively
correlated disorder \cite{AB1,AB2} was considered recently by
Dickman \cite{dickman1} and by Evron et. al. \cite{galit} with
somewhat different  results: while \cite{dickman1} reported that the
critical exponents differ from those of DP transition, in
\cite{galit} it was suggested that only the \emph{off-transition}
behavior of the system is governed by local adaptation to favored
regions, since the scaling function describing the behavior away
from the transition shows significant deviations from the known DP
behavior.

In practice, however, it is hard to avoid quenched heterogeneities
from the substrate on which the process takes place, and this in
true in particular for metapopulations~\cite{hansk} or the spread of
an epidemic \cite{vas, yossibz1}. It turns out that quenched (time
independent) disorder is a relevant perturbation~\cite{jan2} and
seems to change the nature of the transition.

When the system is subject to a global disorder (the spatial range
in which the  heterogeneity occurs is not compact)  a Griffiths
phase exists between the active  and the inactive regions
\cite{mor}. In the parameter region that corresponds to the
Griffiths phase the survival of an occupied patch depends on the
local properties of the substrate, \emph{not} on activation by
neighboring regions. In particular, for each time scale the process
stays alive due to the spatial domains that admit high carrying
capacity~\cite{mor,ot}. Although stochastic fluctuations guarantee
extinction for any localized active patch in the absence of
migration \cite{WKB,meerson,meerson1}, the time to extinction grows
exponentially with the carrying capacity of the spatial domain. This
implies that exponentially rare spatial patches, with high birth
rate, support the population for exponentially long times. An
optimization argument~\cite{mor} shows that in such a case the
survival until $t$ is dominated by rare spatial fluctuations of
linear size $L \sim log(t)$; accordingly, the density falls
algebraically with time. The Griffiths phase is located between an
extinction region, where essentially  no oases exist, and the active
phase, where good oases infect each other to yield a never dying
process.

In this paper we intend to focus on a much simpler problem: the
contact process in the presence of a localized (compact) spatial
heterogeneity, like an oasis in the desert or an ecological trap
(sink habitat). When the bulk system is off criticality it is
segregated into effective patches of size $\xi$, and the effect of a
compact inhomogeneity extends only to this distance, so it cannot
change the long-time properties of an infinite sample. The situation
changes when the bulk system is at criticality: in that case the
correlation length decay obeys a power-law, and in principle a
single oasis (or sink) may affect the properties of the bulk. Here
we present numerical evidences suggesting that   this is indeed the
case, at least in one dimension. A similar phenomenon characterizes
the physics of the zero temperature Kondo effect, when a single
magnetic impurity changes the behavior of an infinite system as it
dictated the density of states close to the Fermi level where
correlation length diverges.

Before proceeding to the results, let us describe the numerical
procedure. We have simulated a  spatial version of the  SIS model
for epidemics, which is a simple generalization of the contact
process, utilizing the  simulation technique presented in
\cite{ks1,yossibz1}. On each site there are $N$ agents, from which
$N-I$ are susceptible and $I$ are infected. If the infection rate is
$\alpha$ and the recovery rate is $\beta$, the chance of an infected
agent to recover within $\Delta t$ is $\beta \Delta t$. On a one
dimensional lattice a susceptible agent on the n-th site has $I_n$
infected neighbors in its own site and $I_{n \pm 1}$ in the
neighboring sites.  If the chance to infect a local susceptible is
$\alpha (1- \chi)/N$ and the chance to infect a neighbor susceptible
is $\alpha \chi/(2N)$, the average number of infections on the
$n$-th site within $\Delta t$ is given by $q_1 (N-I_n)$ where $ q_1
= \Delta t \alpha  \left[ (1- \chi) I_n + \chi I_{n+1}/2 + \chi
I_{n-1}/2 \right]$ and the average number of recoveries is $q_2 I_n$
where $q_2 =  \Delta t \beta I_n$. Once these numbers were
calculated for all the chain, the system is updated by drawing, for
any site,  two numbers from a binomial distribution, one with an
average $q_1 (N-I-n)$ and the other with an average $q_2 I_n$.
$I_n(t)$ is than updated to $I_n(t+\Delta t) = I_n(t) + B(q1, N-I_n)
+ B(q2, I_n)$. This procedure may be considered as a stochastic
Euler integration of the system dynamics, and it converges to the
pure contact process in the limit $N=1, \ \chi = 1/2$. Without loss
of generality we have taken $\beta = 1$ in all cases considered
below, and the generalization to higher dimensions is trivial.

\section{Defect dependent decay exponent in one dimension }

Let us start presenting our results for the  contact process, $N=1$.
The  critical birth (infection) rate in that case is known to be
$\alpha_c = 3.2978$. For a homogenous system at criticality the
survival probability, i.e., the chance  of a cluster, grown from a
single seed, to be active after $t$ steps, is given by $P(t) =
t^{-\delta}$, where the  1d  value is $\delta_0 \approx 0.159$.

\begin{figure}[h]
\begin{center}
\includegraphics[width=0.75\textwidth]{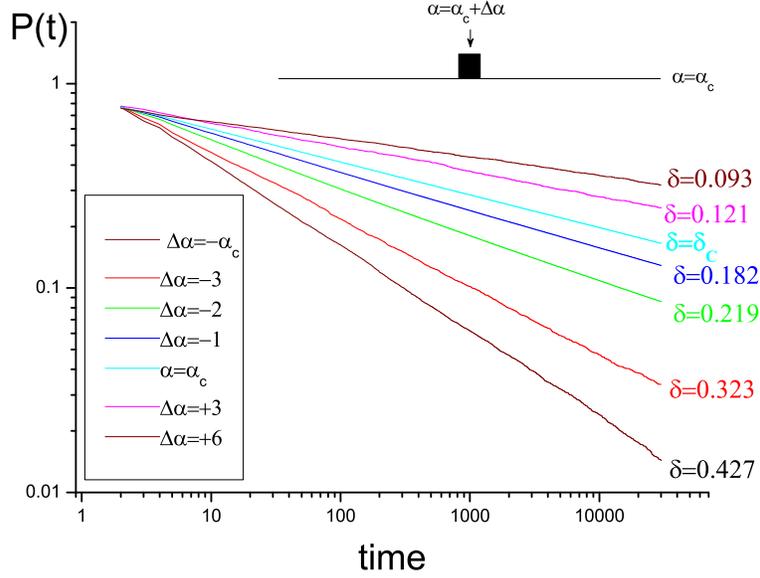}
\caption{The survival probability $P(t)$ vs. $t$ on a double
logarithmic plot for a system of L=5000 sites with periodic boundary
conditions. Only a single particle is allowed on each site, so the
process is equivalent to the contact process. At the bulk the
infection rate is $\alpha_c = 3.2978$, and only on the "exceptional"
site at the origin $\alpha = \alpha_c + \Delta \alpha$. When the
system is homogenous the slope is $\delta_0 \sim 0.16$, as expected
from the theory. If on the central site $\alpha < \alpha_c$ (sink)
the slope is larger, and for an oasis ($\alpha > \alpha_c$) it is
smaller. The results here reflect an average over $10^6$ processes,
all ignited with a single agent on the source/sink
site.}\label{fig1}
\end{center}
\end{figure}

\begin{figure}[h]
\begin{center}
\includegraphics[width=0.75\textwidth]{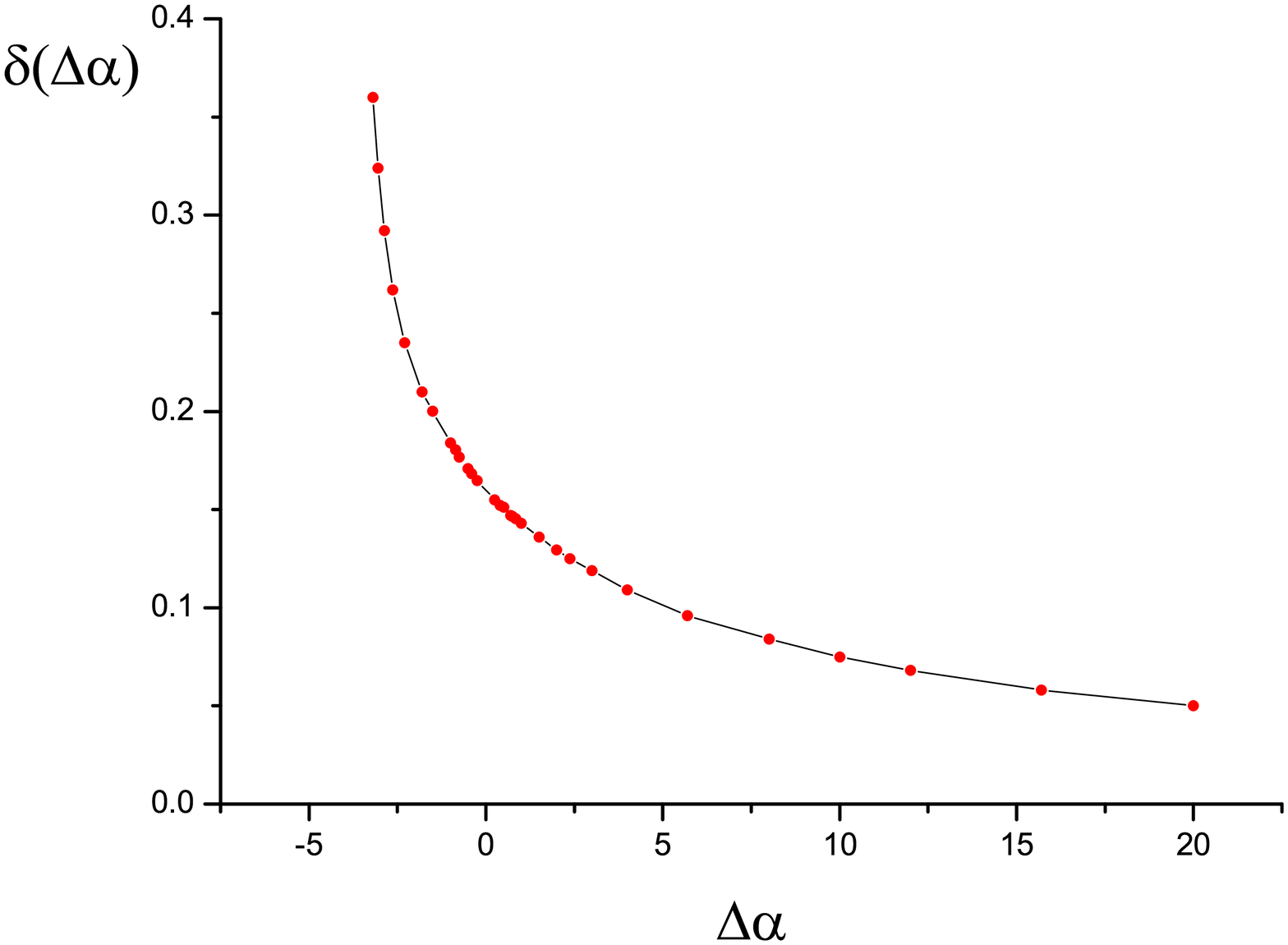}
\caption{The decay exponent $\delta(\Delta \alpha)$ vs. the strength
of the heterogeneity $\Delta \alpha$, from a  sink where the birth
rate is zero ($\Delta \alpha = -\alpha_c$, i.e.,  $\alpha = 0$) to a
very strong oasis.} \label{fig2}
\end{center}
\end{figure}

Figure \ref{fig1} shows what happens when a single site for which
$\alpha = \alpha_c + \delta \alpha$ has been introduced into a
system which is otherwise at criticality (see schematic sketch in
the upper-right part of the figure). When the process is ignited
with a single particle at the heterogeneity (arrow), $P(t)$ is still
a straight line on a double logarithmic plot, but the slope depends
on $\Delta \alpha$: for source (oasis) the slope is smaller than
$\delta_0$  and for sink the decay is faster. This indicate the
failure of the critical system to decouple from  the defect when the
correlation length diverges. Figure \ref{fig2} shows $\delta (\Delta
\alpha)/\delta_0$ between a perfect sink and a very strong source.

What happens when the process is ignited  away from the
heterogeneity? If the oasis/sink is located at the origin and the
process starts when a seed (a single infected person) is positioned
at $x$,  $P(t)$ must decay with the exponent $\delta_0$ until the
system starts to "feel" the oasis, then cross over to the
appropriate value of $\delta$. When the system is at criticality the
spread of a perturbation is controlled by the ratio between the
spatial and the temporal correlation length, i.e.,  a perturbation
is traveling a distance $x$ within a time $t \sim x^z $, where  for
a 1d DP transition $z = 1.58$. This implies that after a time $t$
the source/sink affects a distance $x \sim t^{0.63}$.

\begin{figure}[h]
\begin{center}
$\begin{array}{cc}
\includegraphics[width=3in]{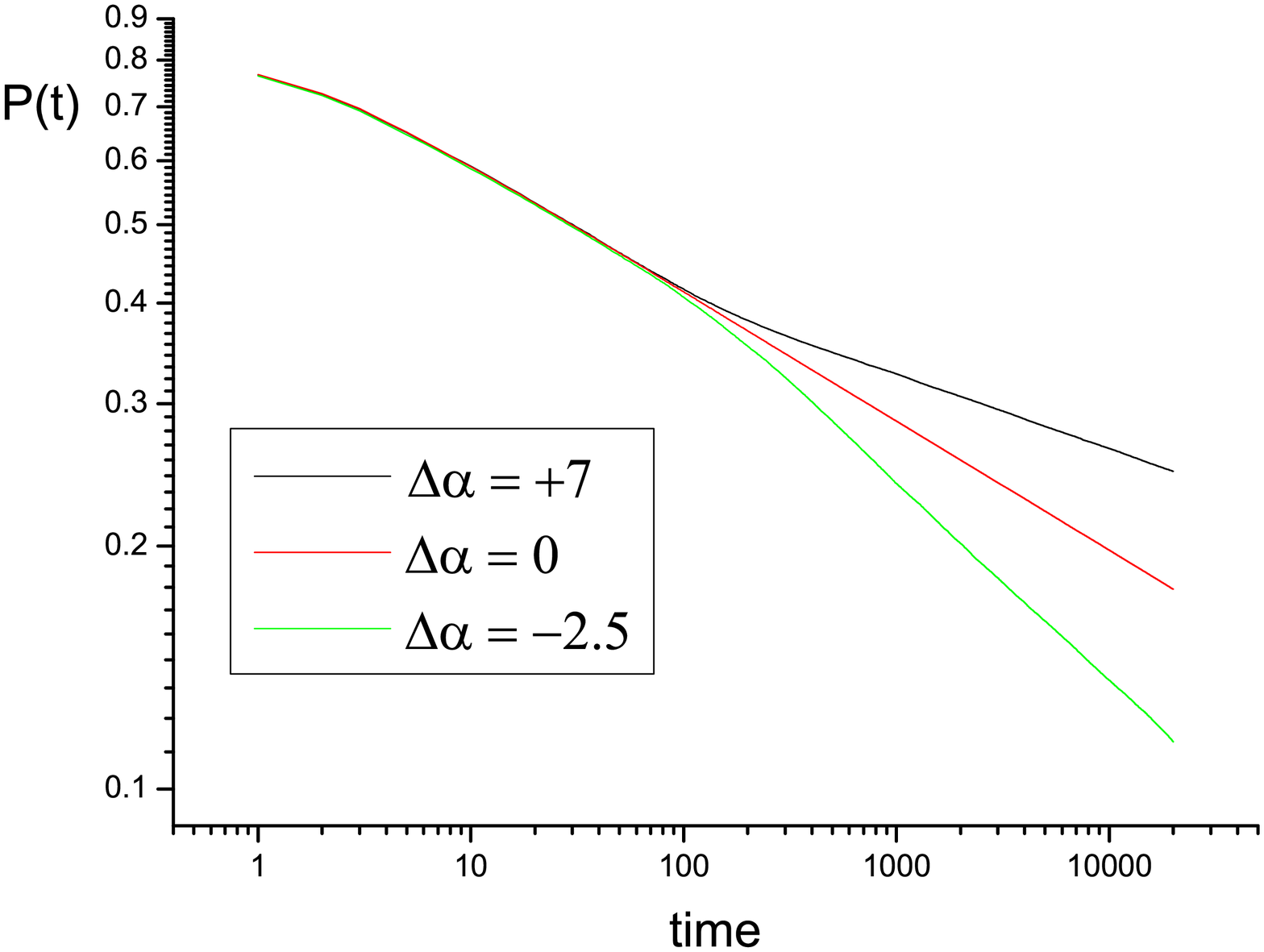} &
\includegraphics[width=3in]{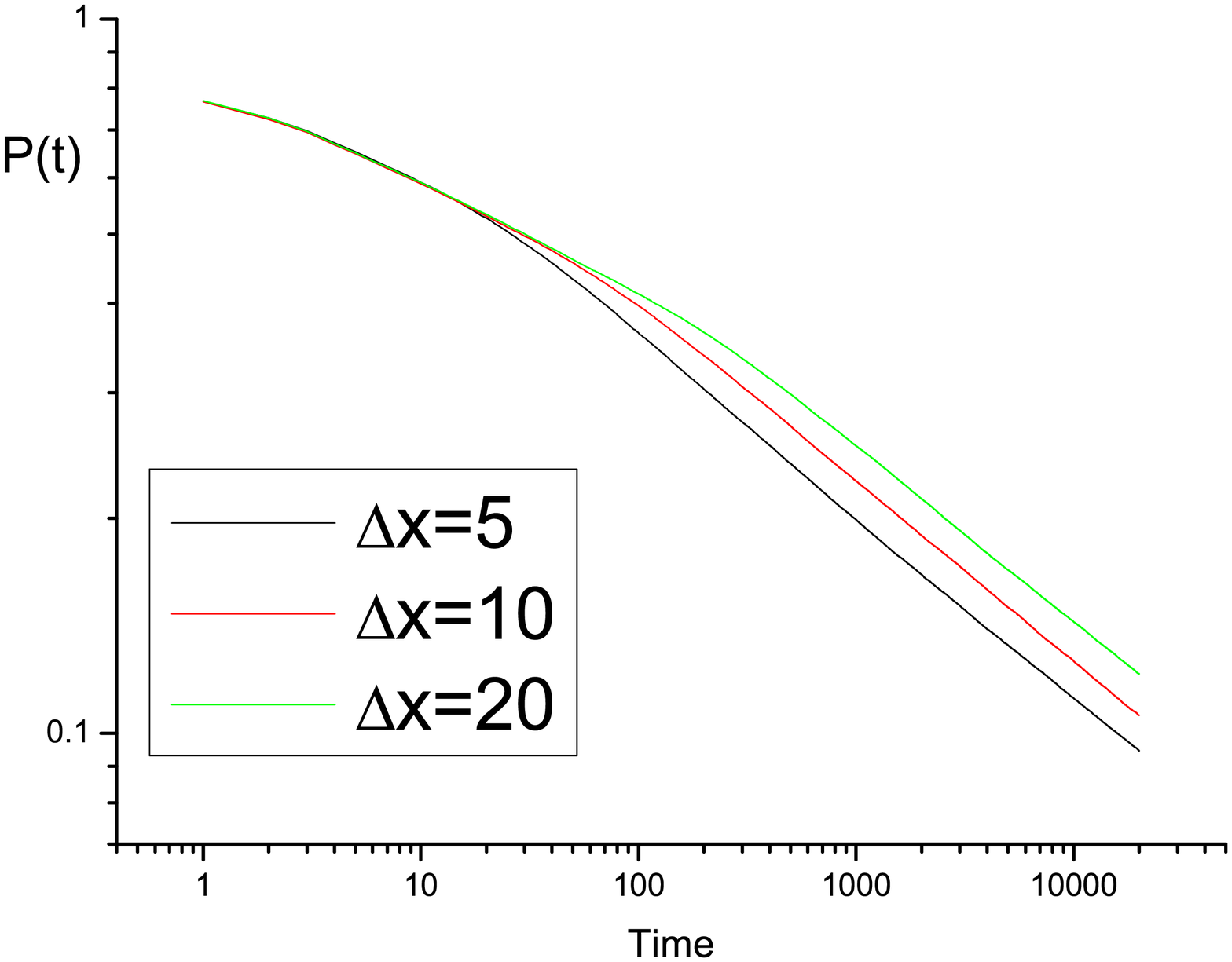}
\end{array}$
\caption{Left  panel: $P(t)$ vs. t (loglog scale) for a contact
process that starts with one agent  at a distance $\Delta x = 15$
from the heterogenous site. Three cases are compared: an oasis
($\Delta \alpha = +7$), a sink ($\Delta \alpha = -2.5$) and a
homogenous system. Right panel: same, with a sink  ($\Delta \alpha =
-2.5$) at the origin, where the three lines correspond to  different
$\Delta x$. }\label{fig3}
\end{center}
\end{figure}

Figure \ref{fig3} illustrates the effect of the distance from the
transition point. Here the seed was deposited at $\Delta x = 15$
where the origin is an oasis, a sink  or a regular site for which
$\alpha = \alpha_c$. $P(t)$ for all cases is identical until the
system start to get some information about the structure at the
origin, than it splits into three different slopes (left panel). In
the right panel there is a sink at the origin and  $P(t)$ is shown
for different distances of the seed from the heterogeneity: one sees
that the short time decay has the exponent of a homogenous system,
the  long time decay also have the same slope, but the crossover
time depends on the distance.

\begin{figure}[h]
\begin{center}
\includegraphics[width=0.75\textwidth]{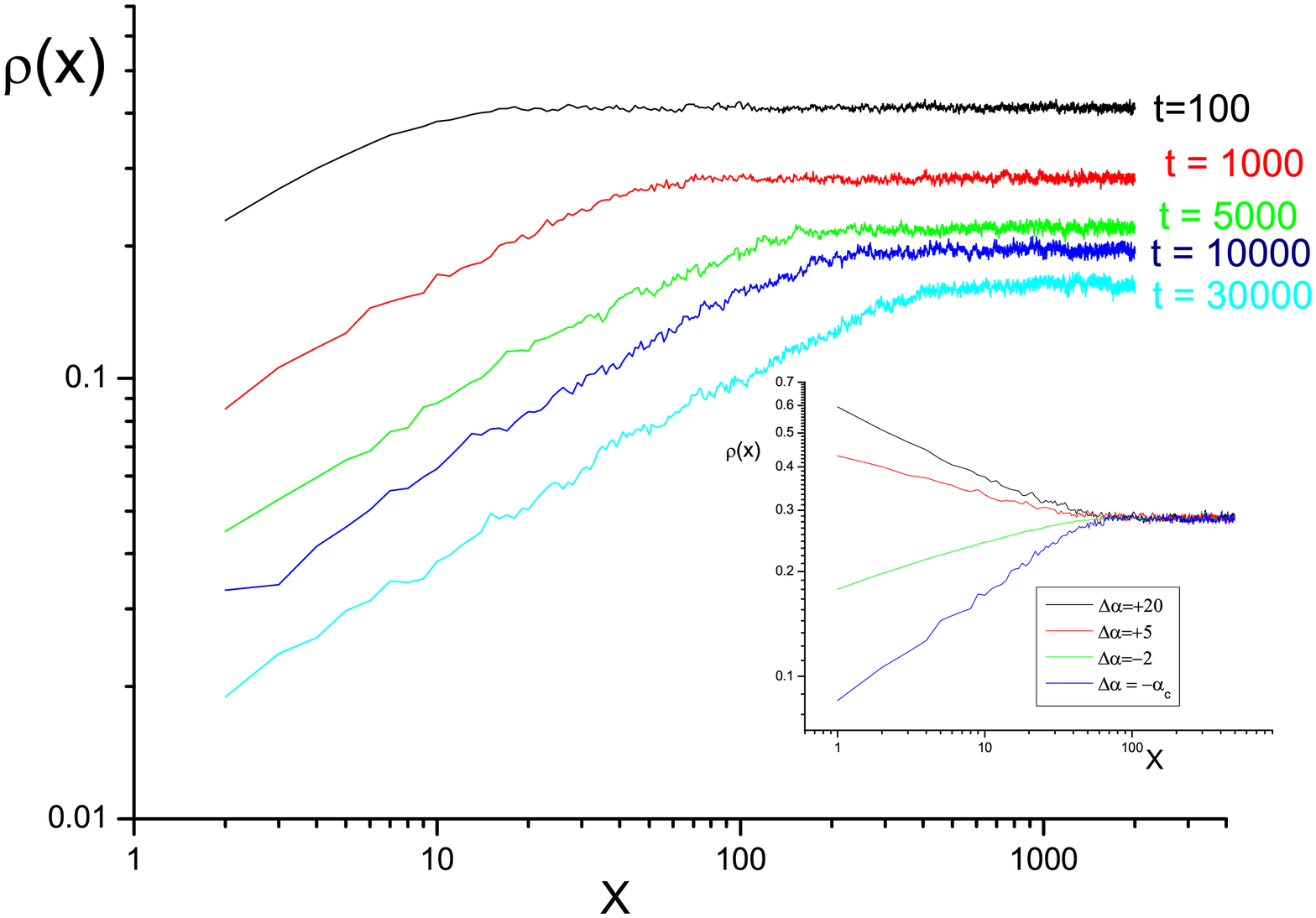}
\caption{The main panel shows a few "snapshots" of the spatial
density profile, $\rho(x)$, for a contact process that starts from a
fully occupied lattice with a sink ($\Delta \alpha = - \alpha_c$) at
the origin. Each line reflects the average results of $1000$
histories. While in the far region the density decays algebraically
with $\delta_0$, in the close region the decay is faster. The region
affected by the sink is also growing. The inset shows the density
profile at fixed time, $t=1000$, for different types of
heterogeneity at the origin. Clearly, in the far region there is no
difference, and the distance in which the sink/source affects the
system is independent of its strength and character. }\label{fig4}
\end{center}
\end{figure}

The same behavior is depicted  in Fig. \ref{fig4}. Here the local
particle density $\rho(x,t)$, starting from a fully occupied
lattice, is shown: a snapshot at $t=1000$ for different types of
special site (inset) and consecutive time frames for a system with a
sink. Since at the close region $\rho(t)$ decays like
$t^{-\delta(\Delta \alpha)}$ and in the far region $\rho(t) \sim
t^{-\delta_0}$, on a double logarithmic scale a straight line
appears for $\rho(x)$ in the intermediate region, with a  slope
\begin{equation}
s = \frac{log(\rho_0 t^{-\delta(\Delta \alpha)}) - log(\rho_0
t^{-\delta_0})}{log(t^{z/2})} \sim 2 \frac{\delta(\Delta \alpha) -
\delta_0}{z}.
\end{equation}
This slope agrees with the measurements from our numerical
experiments.

\section{Universal response to local heterogeneity}

\begin{figure}[h]
\begin{center}
\includegraphics[width=0.75\textwidth]{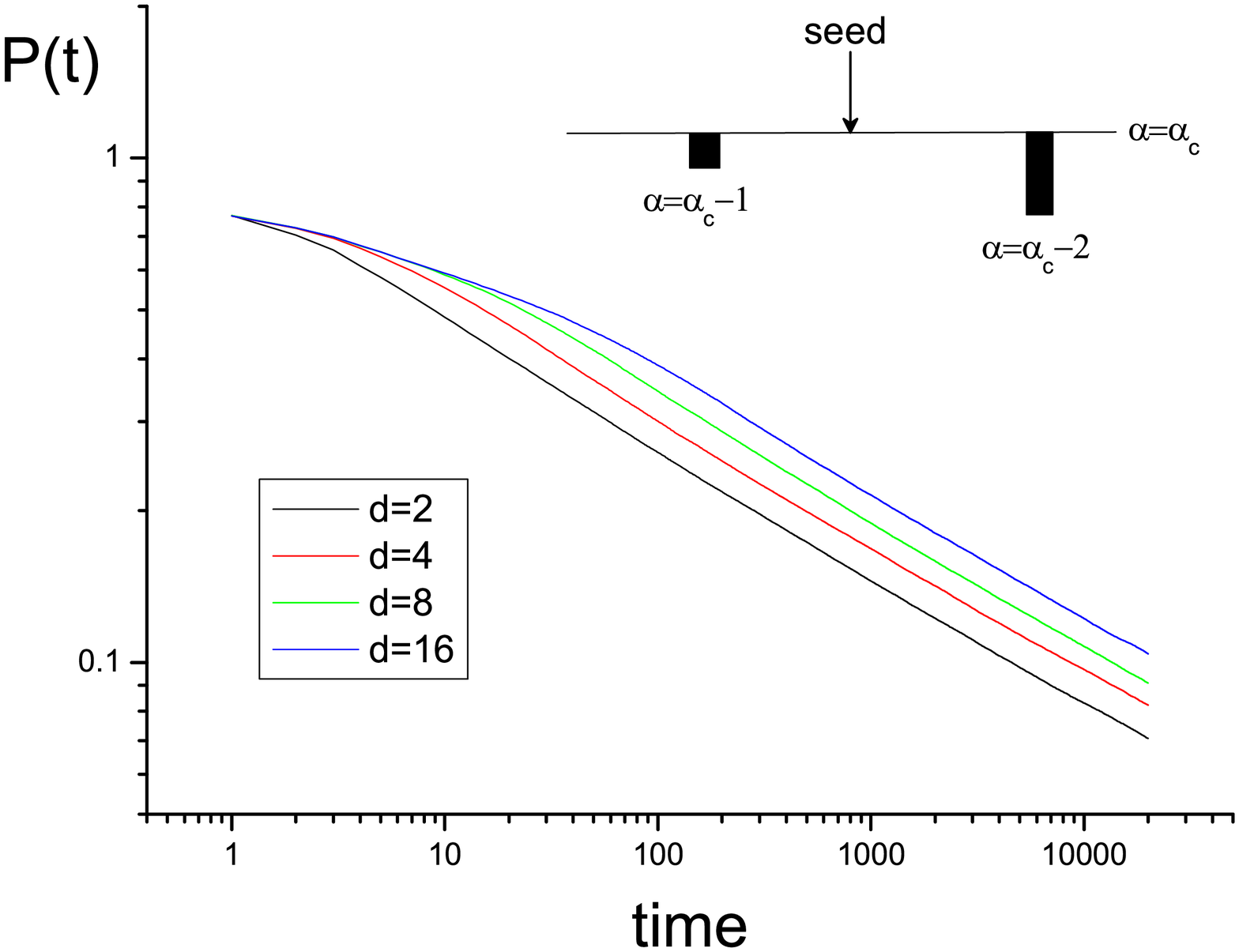}
\caption{The survival probability $P(t)$ vs. t for a contact
process. Here the bulk is at criticality with  two, unequal, sink
sites, one with $\alpha = \alpha_c - 1$ and for the other $\alpha =
\alpha_c - 2$. The process starts when a single seed is introduced
at the origin, which is the midpoint between the two sinks, with a
distance $d$ from the seed to any of the heterogeneities. As $d$
increases, the time it takes to the process to interact with the
sinks is growing, still the asymptotic behavior is the same (the
slope of all these lines is the same within the measurement error).
}\label{fig5}
\end{center}
\end{figure}

Surprisingly, our numerics suggest that the effect of a localized
heterogeneity is also \emph{universal}, i.e., it is independent of
the spatial structure of the source/sink and of $N$, the carrying
capacity of a site. This feature is demonstrated in Figure
\ref{fig5}: here the heterogeneity is made of two exceptional sites,
and the distance between the two, $2d$,  is a  free parameter. When
the process starts with a seed between the two sites, the short time
decay of $P(t)$ is, of course,   nonuniversal.  However, after the
process "probes" the two sites  the long time behavior is universal
and independent of the distance between the sinks. We have carried
out different numerical experiments, with two or three, equal or
unequal, exceptional sites (and for  two sources, source and sink,
two sinks with different $\alpha$s etc.).  In all of them  the tail
exponent is independent of the spatial structure of the compact
heterogeneity.

\begin{figure}[h]
\begin{center}
\includegraphics[width=0.75\textwidth]{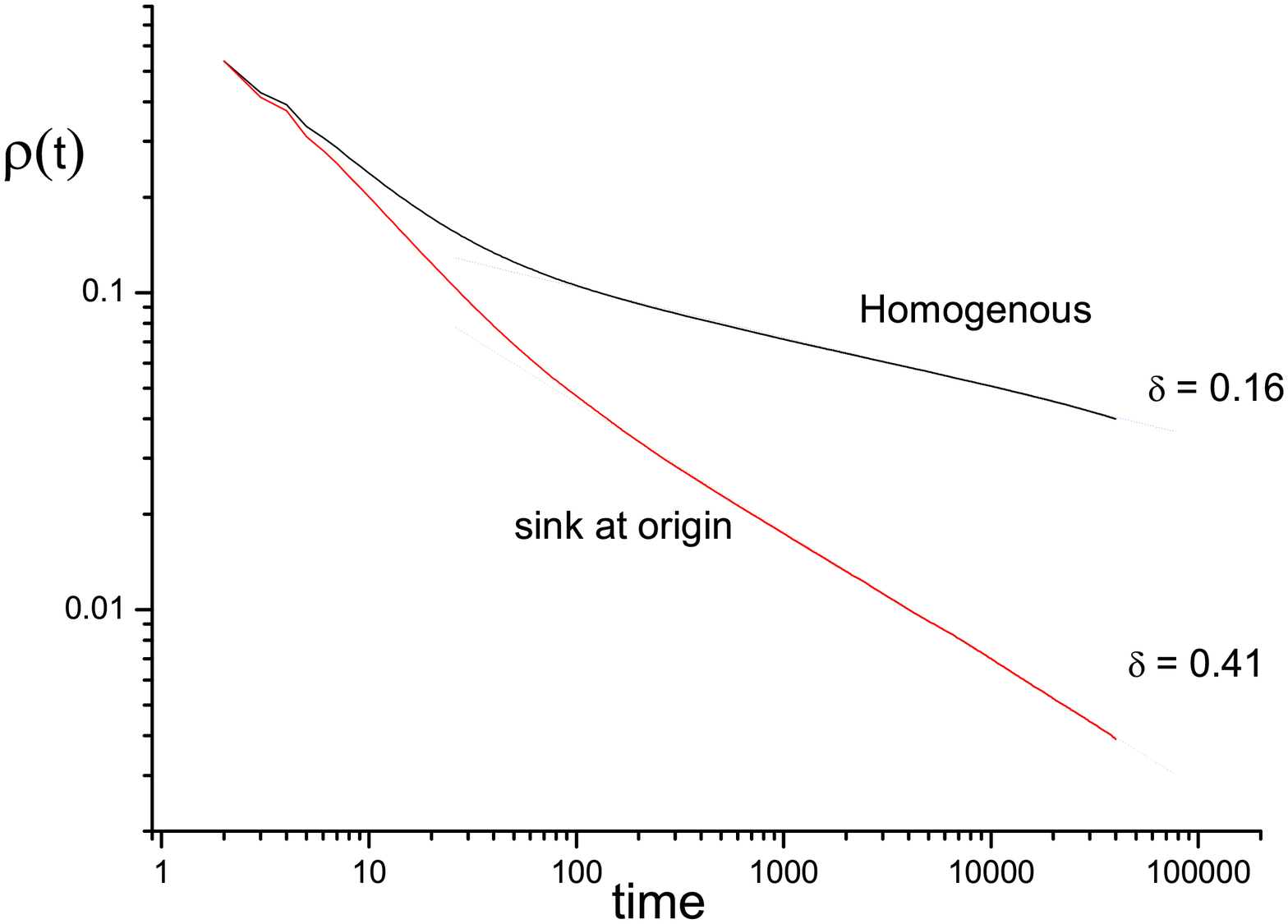}
\caption{The density $\rho(t)$ vs. time for a system with carrying
capacity of 50 agents per site. With $\chi = 0.2$, the transition
for the  homogenous system has been identified at $\alpha_c =
1.1875$. At criticality  (black line) the decay starts with exponent
$\delta = 1$, as expected from the mean-field equations, then when
the stochasticity becomes important it crosses over to $\delta_0$ of
the DP transition (the dashed line is the linear fit for the tail).
On the other hand, if a sink (site with $\alpha = 0$) is introduced
at the  origin the long-time exponent is larger (red line) and the
best fit to the tail (dashed) yields $\delta = 0.41$. The results
reflect an average over $10^6$ histories.  }\label{fig5a}
\end{center}
\end{figure}

The same observation holds if one allows more than one agent at any
site. In Fig. \ref{fig5a} $P(t)$ is plotted for $N=50$. There is a
transient region in which the system followed the deterministic
decay rate $P \sim 1/t$ (as suggested by the equation $\dot{I} =
\alpha \chi \nabla^2 I - \alpha I^2$, which is the deterministic
limit of the stochastic process defined above~\cite{ks1}). However,
beyond this transient region the stochastic exponent controls the
decay, with different exponents for the homogenous and the
inhomogeneous case. The values of the asymptotic decay $\delta$
seems to be very close for $N=1$, $N=10$ (not shown) and $N=50$
(differences smaller than 0.02, which is more or less the numerical
error in the measurement of the slope), so it seems that  the
exponent is independent of $N$ either.

\section{Two and three spatial dimensions: convergence to the
homogeneous critical behavior}

\begin{figure}[h]
\begin{center}
$\begin{array}{cc}
\includegraphics[width=3.5in]{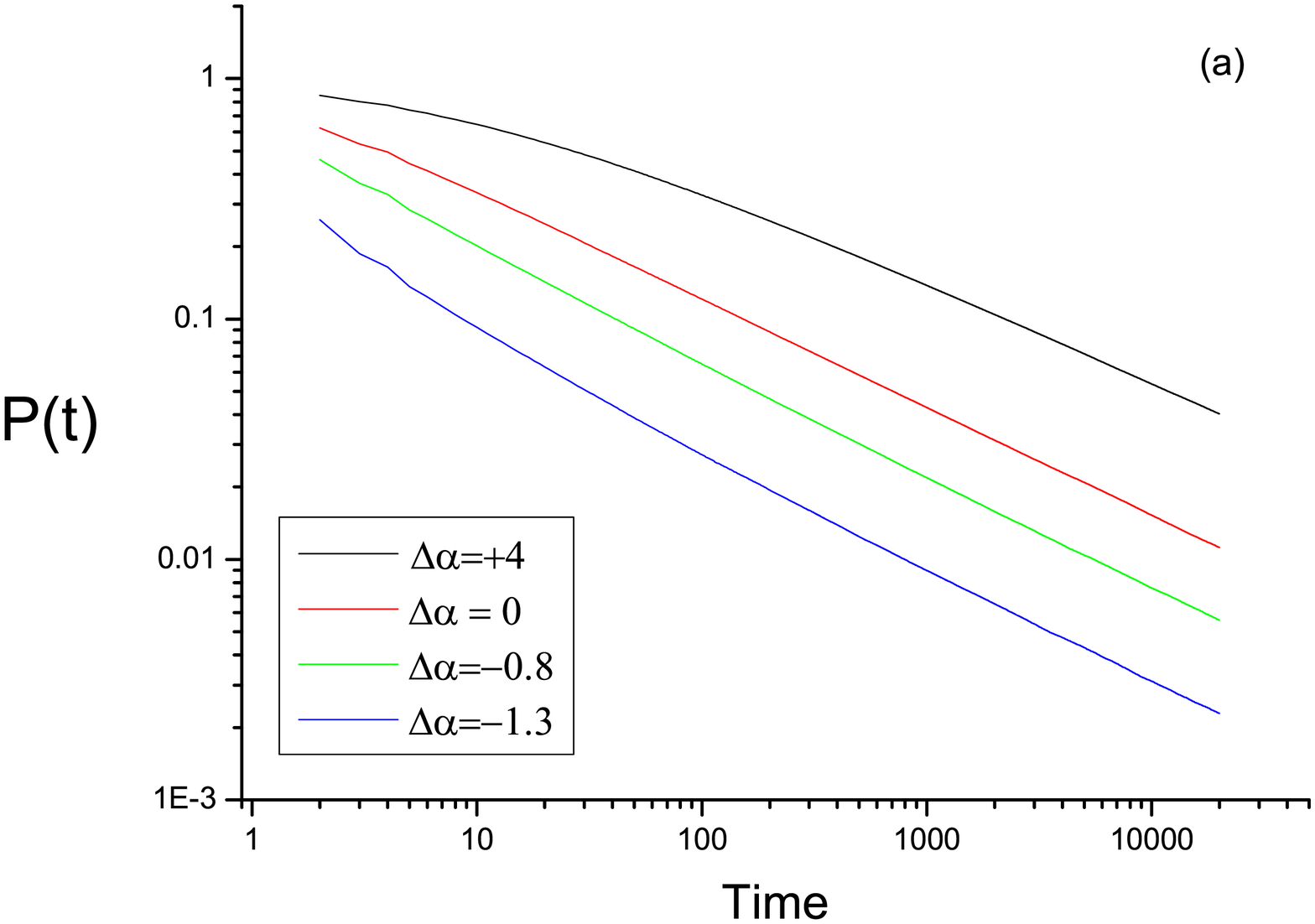} &
\includegraphics[width=3.5in]{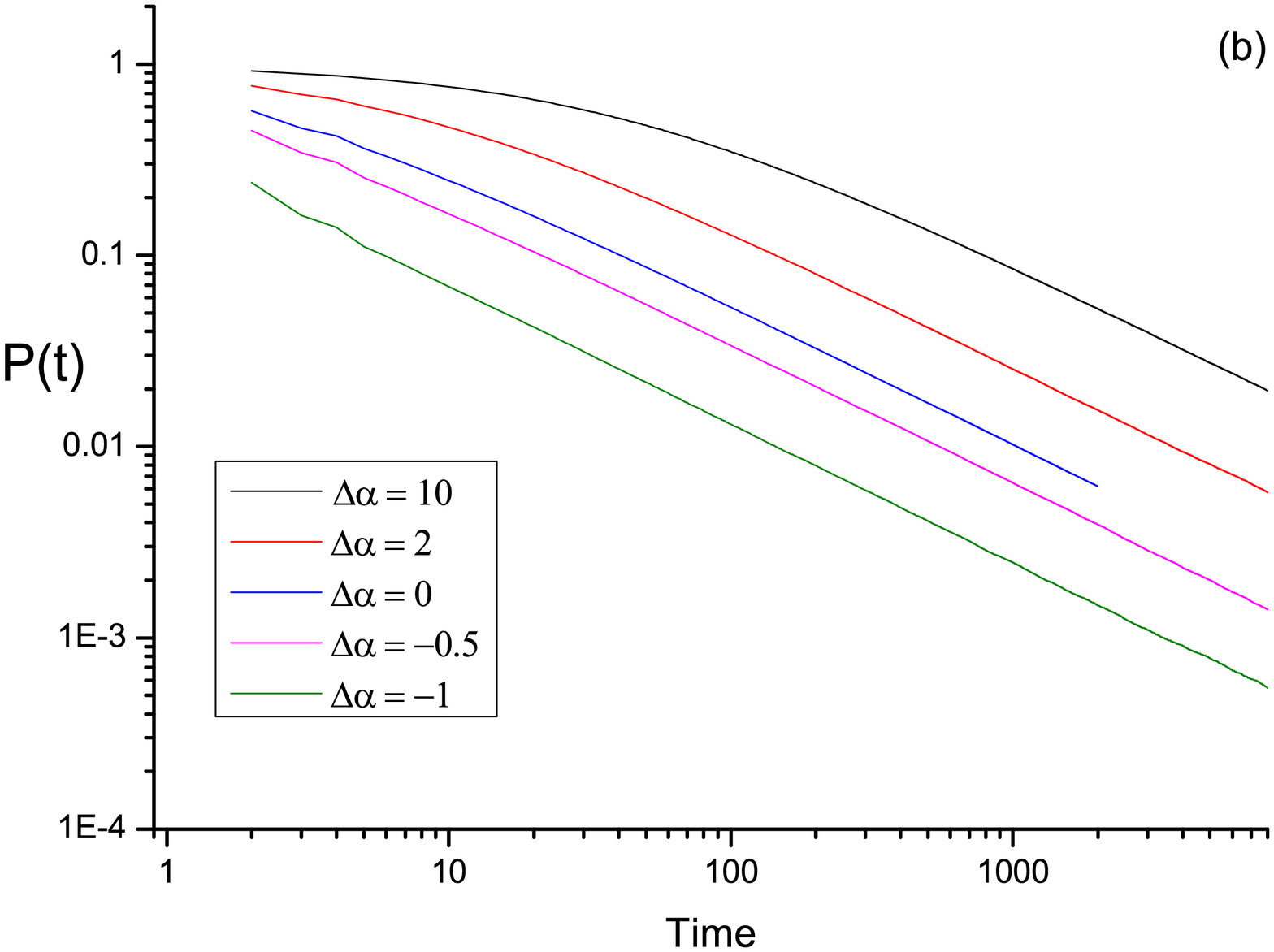}
\end{array}$
\caption{The survival probability for a process initiated with a
single seed at the origin for 2d (a) and 3d (b). The site at the
origin has different reproduction rate. Either for source or for
sink, the long-time decay seems identical and converge to the known
values for the DP exponents in these dimensions. The bulk values of
$\alpha_c$ are $1.6488$ for 2d and $1.3168$ for 3d. }\label{fig6}
\end{center}
\end{figure}

All the results presented so far are for a 1d system. In higher
dimensions the situation is different: here the effect of a compact
heterogenous source/sink manifests  itself only as a transient, but
the asymptotic behavior of $P(t)$ follows the exponent $\delta_0$
that characterizes the homogenous system, as shown in Figures
\ref{fig6}. Careful examination of the results seems to indicate
that the long time decay is a combination of two exponent of the
form:
\begin{equation}
P(t) \sim t^{-\delta_0} \left(1 + A t^{-\delta_1} \right)
\end{equation}
for a sink and
\begin{equation} \label{33}
P(t) \sim  \frac{t^{-\delta_0}}{ \left(1 + A t^{-\delta_2} \right)}
\end{equation}
for a source, i.e., that the transient behavior itself decays
algebraically in time, with an exponent larger than $\delta_0$.

The qualitative difference between the behavior in 1d and in higher
dimension suggest that the effect of a single defect has to do with
the chance of a random walker to return to the origin. At
criticality the average spatial size of the contact process is not
shrinking or growing, and since there is no preferred direction the
process visits spatial points at random (this feature has been
already considered in the theory of persistence time
\cite{persistence}). Since in above 2d there is a good chance that
the process never returns to the oasis, and 2d is marginal, only in
1d the oasis changes the critical exponent.

\section{Discussion}

Along this paper we have dealt with a single oasis/sink coupled to a
critical bulk. This problem is quite similar to the one recently
considered by Barato and Hinrichsen~\cite{hin1}, in which an inert
spatial system (at the bulk there are no birth-death processes, only
diffusion) is coupled to a single active site on the boundary. In
this setup the active site admits only one agent ($N=1$) with and is
"critical" in the sense that the birth rate on this site is equal to
the death rate. These authors reported a decay exponent $1/6$  for
that system. Later, Burov and Kessler~\cite{stas} have solved
analytically the problem for $N=\infty$ and found  $\delta = 1/4$;
For any finite $N$ their numerics indicated algebraic decay with
different exponent.

In some sense, the model considered here and the model of
\cite{hin1,stas} are similar, as they both deal with an exceptional
site coupled to a critical bulk. The main difference is that in our
model the bulk is active and the criticality is "quantum", while the
model considered before has an inert bulk. Still it will be
interesting to check the response of the inert  bulk model to a
spatially structured defect.

An interesting question that may be asked in general with regard to
these systems  is about their deterministic limit, i.e., the
behavior when $N \to \infty$. For the deterministic system the
critical point is at $\alpha = 1$ and a single defect coupled to a
critical medium satisfies an equation like  $$\dot{I} =  \nabla^2 I
+ \alpha(x) I -  I^2,$$ where $\alpha(x)$ is takes nonzero values at
the spatial region associated with the defect. Since any potential
well in 1d supports a localized state, so does the linearized
evolution operator ${\cal L} = \nabla^2 + \alpha(x)$. Thus for any
oasis, with any spatial structure, the system must admit a localized
colony (its amplitude is dictated by the nonlinear term) that lives
forever~\cite{ns}. $P(t,N)$ is thus singular at the deterministic
limit. In higher dimensions, on the other hand, not any oasis
supports a localized state, so the deterministic limit may be
non-singular.

\vspace*{.4in}

%\bibliographystyle{pnas}
%\bibliography{bibfile}

\end{document}